\documentclass[twocolumn,twoside,slac_two]{revtex4}
\usepackage{graphicx}
\usepackage{fancyhdr}
\usepackage{hyperref}
\hypersetup{
    colorlinks=true,
    urlcolor=magenta,
}
\begin{document}

%Title of paper
\title{Facilities of Athens Neutron Monitor Station to Space Weather services}

\author{H. Mavromichalaki, M. Gerontidou, P. Paschalis, E. Paouris}
\affiliation{Athens Cosmic Ray Group, Faculty of Physics, National and Kapodistrian University of Athens, 15784 Athens, Greece}

\begin{abstract}
In the frame of the comprehensive knowledge, detection and forecasting of the solar –terrestrial relations as well as space weather events, the ground based measurements of the network of neutron monitor constitutes a vital tool for these studies.  This is mainly the reason that Athens Neutron Monitor Station (A.Ne.Mo.S.)  beyond of the provision of its real time data, has also developed several research applications. More specifically, applications such as a) an optimized automated Ground Level Enhancement Alert (GLE Alert Plus) b) a web interface, providing data from multiple Neutron Monitor stations (Multi-Station tool) and c) a simulation model, named DYnamic Atmospheric Shower Tracking Interactive Model Application (DYASTIMA), which allows the study of the cosmic ray showers resulted when primary cosmic ray particles enters the atmosphere, have been developed. The two first applications are currently federated products in European Space Agency (ESA) and actually available via the Space Weather Portal operated by ESA. On the other hand, the  contribution of the simulation tool DYASTIMA, based on the well known Geant4 toolkit, to the calculations of the radiation dose received by air crews and passengers within the Earth's atmosphere led us to develop an extended application of DYASTIMA named DYASTIMA-R. This new application calculates the energy that is deposited on the phantom and moreover the equivalent dose. Furthermore, a Space Weather Forecasting Center which provides a three day geomagnetic activity report on a daily basis has been set up and is operating at the Athens Neutron Monitor Station.
Finally, all above developed services are in essential importance for the fundamental research as well as for practical applications concerning Space Weather.  

\end{abstract}

\maketitle

\thispagestyle{fancy}

\section{INTRODUCTION}
Ground based cosmic ray detectors measure secondary cosmic rays, i.e. the different components produced in air showers in the Earth's atmosphere. In particular the Neutron monitor detectors  measure the nucleonic (protons and neutrons)  component and their  recordings have  been used since 1950's. Despite their decades of tradition, ground based neutron monitors (NMs) remain the state-of-the-art instrumentation for measuring cosmic rays, and they play a key role as a research tool in the field of space physics, solar-terrestrial relations, and space weather applications. They are sensitive to cosmic rays penetrating the Earth's atmosphere with energies from about 0.5-20 GeV, i.e. in an energy range that cannot be measured with detectors in space in the same simple, inexpensive, and statistically accurate way \cite{Simpson 2000}.

\begin{figure}[t]
\centering
\includegraphics[width=75mm]{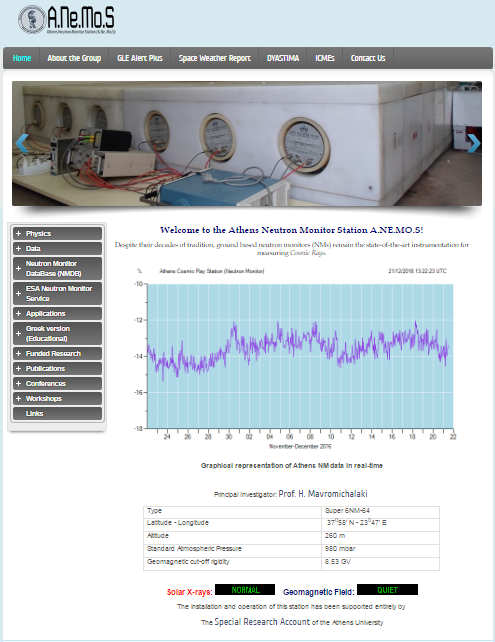}
\caption{Athens Neutron Monitor Station - A.Ne.Mo.S. website.\label{Anemos}}
\end{figure}

Since 2000 the Athens Neutron Monitor Station (A.Ne.Mo.S.) (6NM-64) provides cosmic ray data in real time with minimum resolution of 1 minute. The measurements can be obtained via the internet in several formats and resolutions (Figure \ref{Anemos})  (http://cosray.phys.uoa.gr/).

Since 2008 a database collecting the high resolution measurements of neutron monitors around the world is implemented, with the participation, at first, of 12 countries and more than 50 Neutron Monitor Stations. A.Ne.Mo.S sends real time data to this European Neutron Monitor Database (NMDB) every 1 minute. A  mirror server of the NMDB database has been set up and is in operation at the Athens NM station. Athens NM database provides cosmic ray corrected data with five different algorithms in graphical and digital forms \cite{Mavromichalaki 2010}, \cite{Mavromichalaki 2011}.

In addition, the barometric coefficient, necessary for the pressure correction of data, can be calculated for every NMDB station via a new online tool. The accuracy of the calculation, expressed via the correlation coefficient of the linear regression, reflects the data quality of the stations.

\section{European Neutron monitor services}
The Athens Cosmic Ray group participates as an expert  group to the European Space Agency (ESA), providing two different services. In the first framework of the SSA Space Radiation Condination Center is the Multi stations which is a web interface with access to the archived neutron monitor data from multiple stations and the user ability of interactivness, i.e. a user may chose to plot on their own a specific time interval with 1-min and/or 1-hour resolution. The second one is a timely and accurate warning for GLE (GLE Alert) based on multiple Neutron Monitor data. 

\begin{figure}[t]
\centering
\includegraphics[width=75mm]{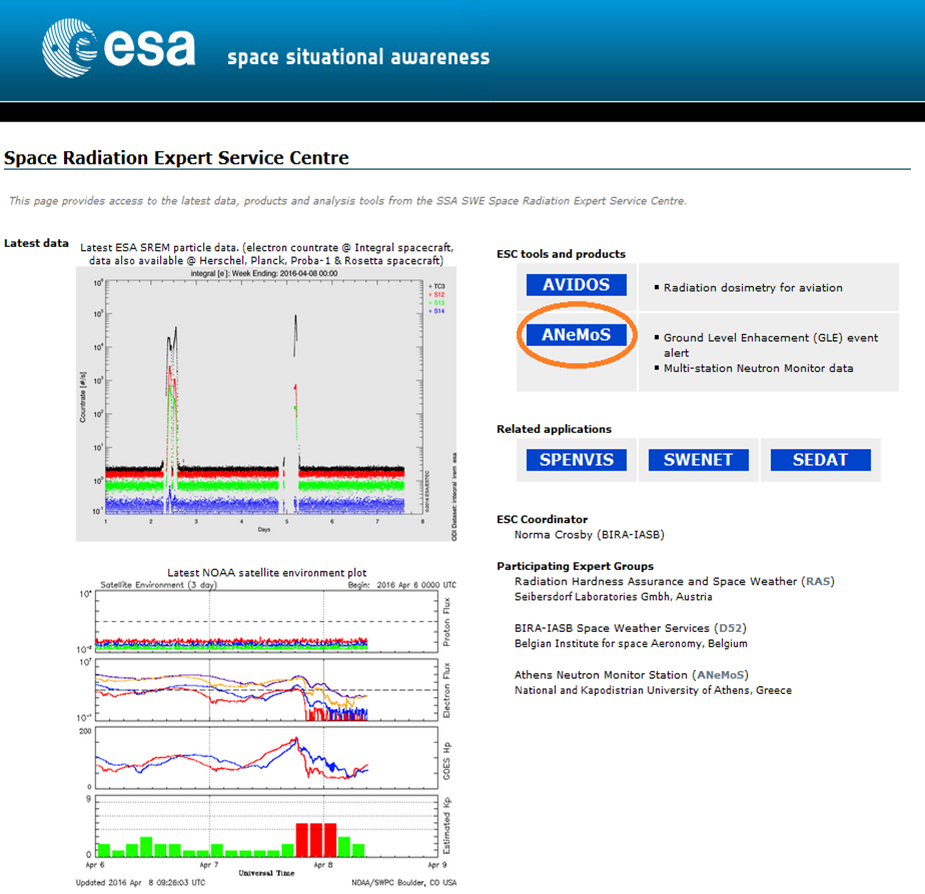}
\caption{The European Space Agency portal.\label{ESA}}
\end{figure}

These services of A.Ne.Mo.S. are provided via the portal of ESA (http://swe.ssa.esa.int/web/ guest/space-radiation), as it is seen in Figure \ref{ESA}. 

\subsection{Multi Station}
 A.Ne.Mo.S. group has established the communication bridge with the Neutron Monitor Community via the Neutron Monitor Database (NMDB) (www.nmdb.eu). The aforementioned service utilizes the mirror server of the NMDB that is in operation at the premises of A.Ne.Mo.S. at National and Kapodistrian University of Athens. Care has been taken in order to facilitate an easy access to end users of the service as well as to cover as complete as possible the potential requests to the service by the users. 
 
 The multi-station interface provides an easy way to access the data that are stored in the NMDB database.  The user has to select the stations, the variables, the time interval and the resolution of the exported data. The output can be obtained in both plot and ascii format. Moreover, a feature that allows the retrieval of data in csv file has been implemented, allowing further processing of the data by the user

\subsection{Ground Level Enhancement Alert- GLE Alert Plus}
This  system requires the availability of as many selected neutron monitors as possible, providing 1 minute resolution data, updated every 1 minute. A local database operated at the A.Ne.Mo.S.. The output is an automated real- time GLE Alert. A graphical web interface denoting the evolution of the GLE Alert and its various pre-Alert stages, i.e. warning stage, has been implemented and provided via link to the SSA center.The recordings of each Neutron Monitor station providing data to the NMDB are the input data for the GLE Alert Plus. For every 1 minute it calculates the moving average of the previous hour (i.e. 60 1-min measurements) and the threshold that represents the upper limit for which each NM station is considered to be at ‘Quiet’ mode. If three consecutive 1-min measurements exceed this threshold, the particular NM station is considered to be at a ‘Station Alert’ mode and an elapsed time window of 15 min is being triggered. In case 3 NM stations, independently of each other, enter the ‘Station Alert’ mode within the aforementioned time window a General ‘GLE Alert’ is being marked and an Alert is issued.

\section{Athens Space Weather Forecasting Center }
From the beginning of the year 2012 a new service named Athens Space Weather Forecasting Center - ASWFC is operated at the Athens Neutron Monitor Station. 
The continuous space measurements by the satellites ACE, SOHO, GOES, SDO, PROBA, STEREO A and B, together with ground based observatories as neutron monitors and magnetometers have led to the implementation of Space Weather Centers for the short and long term forecasting of the planetary geomagnetic index Ap. The ASWFC provides a daily report that includes current geomagnetic conditions in near-Earth space \cite{McPherron 1999}, \cite{Abunina 2013} as well as a 3-day forecast of the planetary geomagnetic index Ap. This estimation of the Ap index is based on a set of rules that includes a number of known parameters/properties of Ap index \cite{Gerontidou 2013}.

\begin{figure}[t]
\centering
\includegraphics[width=75mm]{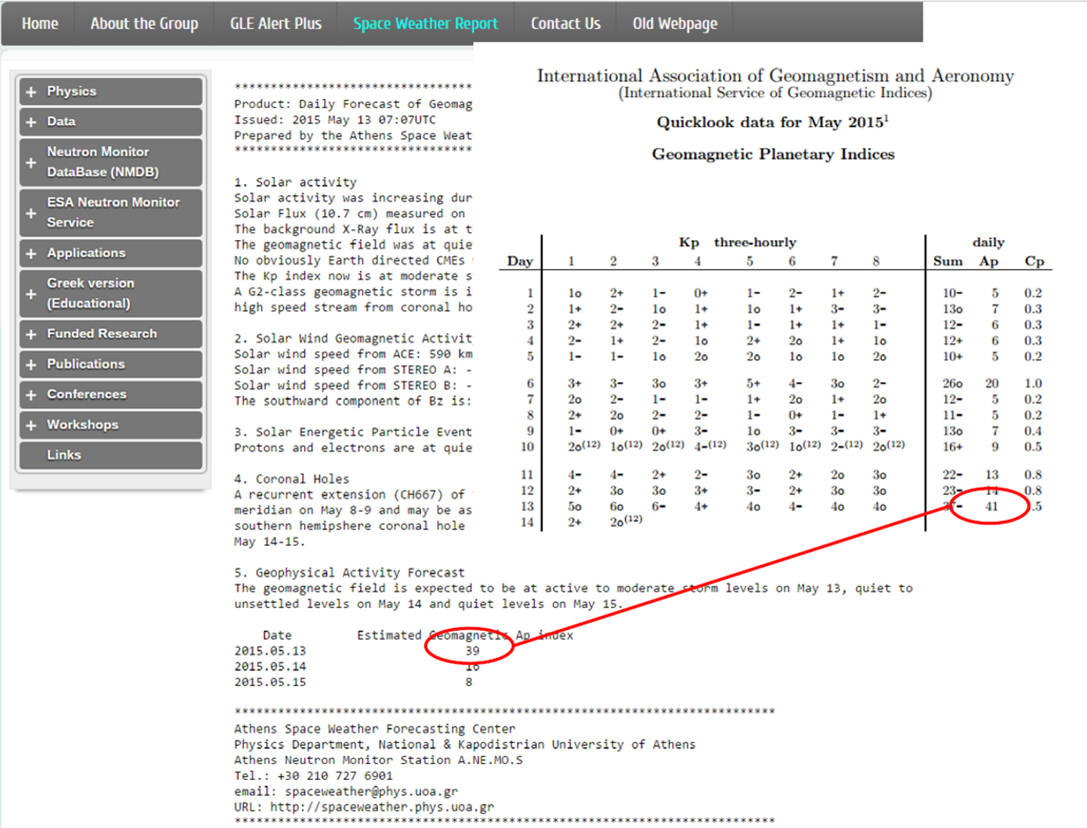}
\caption{A successfull forecasting of 13 May 2015 geomagnetic storm.\label{ASWFC}}
\end{figure}

The  estimation of the Ap index in the Athens forecasting center is based on the current observations of the Sun and near-Earth space from all above satellites covering  a wide range of Sun, an autoregressive model (AR model)  which is taking into account the solar events, CMEs and Coronal holes, magnetic activity 27-days before, the phase of  solar cycle as well as the  yesterday$' $s Ap index. This is due to the fact that Ap index seems to be persistent in the sense that yesterday’s Ap index may be the same today. Moreover, if a trend is being marked in Ap behaviour, i.e., if Ap is being decreasing or in contrast if it has been increasing for several days, this trend may continue, as this may be due to a recurrent behavior of the Sun or the passage of a transient magnetic field (Figure \ref{ASWFC}).

We consider the magnetic activity 27-days before, when due to the solar rotation the same active region is facing the Earth. Also, the phase of the solar cycle is taken into account, i.e., if they provide the forecast in solar minimum conditions, it is most likely for the Sun to produce recurrent behaviour. Finally, consult all available data from the Sun, near-Earth space and the Earth  that may demonstrate signs of intense activity which will result into an increase of the geomagnetic conditions and the Ap index.

\section{The new simulation model DYASTIMA}
A  simulation model, named DYnamic Atmospheric Shower Tracking Interactive  Model Application (DYASTIMA), based on the well known Geant4 toolkit has been developed in A.Ne.Mo.S. This model allows the study of the cosmic ray showers that are developed when primary cosmic ray particles enter to the atmosphere. The neutron monitor response to the several cosmic ray particles that reach the Earth$' $s surface can be determined by using the simulation application of the 6NM-64.\\

\begin{figure}[h]
\centering
\includegraphics[width=75mm, height=70mm]{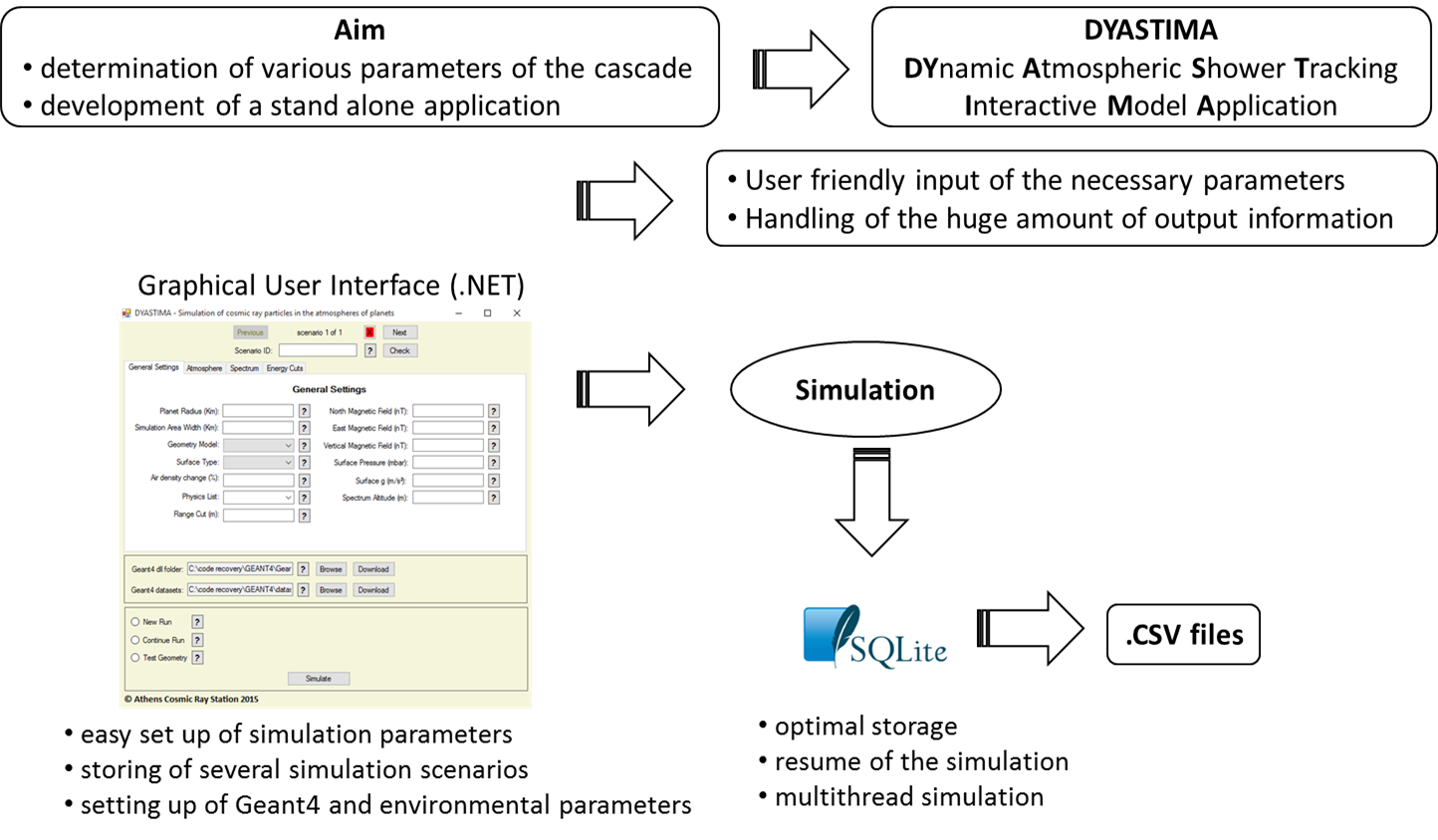}
\caption{DYASTIMA's algorithm.\label{DYASTIMA}}
\end{figure}

DYASTIMA supports the definition of a beam consisting of particles that vary in type, energy and direction \cite{Paschalis 2014}. The output data are exported in a detailed or in a synoptic form, ready to be used for plots, histograms or contour figures (see Figure \ref{DYASTIMA}).
 
A new software application DYASTIMA-R, which constitutes an extension of DYASTIMA uses the output provided by DYASTIMA, in order to calculate the energy that is deposited on the phantom and moreover the equivalent dose \cite{Paschalis 2016}. Monte Carlo simulations are made in order to describe the particle interactions and the transport of the primary and secondary radiation through matter, especially through simulated media, such as the human body (phantom) and the aircraft shielding (optional).

\bigskip 
\begin{acknowledgments}
Athens Neutron Monitor station is supported by the Special Research Account of Athens University (70/4/5803). The research of the  European Neutron Monitor Services has received funding under ESA Tender: RFQ/3-13556/12/D/MRP. We acknowledge the NMDB database (www.nmdb.eu), founded under the European Union's FP7 programme (contract no. 213007) for providing cosmic ray data
\end{acknowledgments}

\end{document}